\documentclass[a4paper,fleqn,usenatbib,referee]{mnras}

\usepackage{mathptmx}

\usepackage[T1]{fontenc}
\usepackage{ae,aecompl}
\usepackage{graphicx}
\usepackage{amsmath}
\usepackage{lscape}
\usepackage{subfigure}
\usepackage{epstopdf}
\usepackage{multirow}
\usepackage{placeins}

\usepackage{graphicx}	
\usepackage{amsmath}	
\usepackage{amssymb}	

\title[The eccentric BEER approximation]{BEER analysis of Kepler and CoRoT light curves. V. eBEER: Extension of the Algorithm to Eccentric Binaries} 

\author[M. Engel et al.]{
M. Engel,\thanks{E-mail: michayesh@gmail.com}
S. Faigler,
S. Shahaf
and T. Mazeh
\\
School of Physics and Astronomy, Raymond and Beverly Sackler Faculty of Exact Sciences, Tel Aviv University, Tel Aviv  69978, Israel
}

\date{Accepted 2020 July 16. Received 2020 July 15; in original form 2019 October 16}

\pubyear{2019}
\def\astrobj#1{#1}
\begin{document}
\label{firstpage}
\pagerange{\pageref{firstpage}--\pageref{lastpage}}
\maketitle

\begin{abstract}
	We present an extension of the BEER model for eccentric binaries -- eBEER, approximating the BEaming, Ellipsoidal and Reflection effects by harmonic series of the Keplerian elements of their orbit.
	As such, it can be a tool for fast processing of light curves for detecting non-eclipsing eccentric binary systems.
	To validate the applicability of the eccentric model and its approximations, we applied eBEER to the {\it Kepler} light curves, identified a sample of bright non-eclipsing binary candidates, and followed three of them with the Wise observatory eShel spectrograph. 
After confirming the three systems are indeed radial-velocity (RV) binaries, we fitted the light curves {\it and} the RV data with PHOEBE, a detailed numerical light-curve and RV model, and showed that the PHOEBE derived parameters are similar to those obtained by the eBEER approximation.
\end{abstract}



 
\section{Introduction}

The {\it Kepler} space mission \citep{Borucki2003} provided more than 100,000 stellar light curves with a time span of 4 years, with unprecedented precision that can get to several tens of ppm.
\cite{LoebGaudi2003} and \cite{Zucker2007} showed that such precision allows the identification of non-eclipsing  binaries and sometimes even planets by detecting the BEaming, Ellipsoidal and Reflection (BEER) effects.

The beaming modulation is due to a relativistic Doppler effect that causes a change in the stellar brightness induced by its orbital radial velocity (RV).
The ellipsoidal modulation is caused by the tidal distortion of the stellar shape which rotates with the binary period.
The stellar radiation reflected by its companion results in the third modulation.  
The combination of the three effects creates a periodic photometric modulation with a specific signature, allowing the discovery of non-eclipsing binary systems. 

\cite{Faigler2011a} developed an algorithm to search for non-eclipsing binaries in the {\it Kepler} and {\it CoRoT} 
\citep{corot06, corot09}
light curves for {\it circular} orbits. They used a rather simple approximate representation of the three-component modulation in terms of sine and cosine of the first two harmonics of the orbital period. For the detected systems, the algorithm yielded an estimation of the masses and radii of the two components. 
This search led to the detection of binaries and even planets, as demonstrated by several studies \citep[e.g.,][]{Faigler2012,Faigler2013,Faigler2015,TalOr2015, Faigler2015b, 
millholand16}.

However, the BEER search is limited to circular-orbit binaries and might fail to detect eccentric binaries that are common, 
especially if their orbital periods are larger than a few days. Many such eccentric binaries were detected and analyzed by previous studies \citep[e.g.,][]{Soszynski04,nicholls12}. 
Similarly, many eccentric exo-planets were found by the RV and  transit techniques, even with extreme eccentricities
\citep[e.g.,][]{naef01}.

In that context, \cite{penoyre19} presented an analytic approximation for the ellipsoidal variability of the host luminosity induced by an eccentric planet.
\cite{penoyre19a} extended this work to model the host luminosity variability due to the three BEER effects induced by an eccentric {\it planet}, and analyzed the prospects for detecting such variability in the TESS \citep{ricker15} light curves and the corresponding power spectra. 
However, these models 
focus on the ellipsoidal leading term in $R/a$, the host radius in semi-major axis units, and  do not model any specific observed-band flux.

This paper presents eBEER, a generalized BEER model for eccentric binary systems, using the Keplerian eccentric motion, while modeling the flux contribution of the two components at the observed band. This is an approximate model that depends on the orbital parameters, the eccentricity and the argument of periastron in particular, which, obviously, do not appear in the circular model.

To validate the applicability of the eccentric model and its approximations, we applied the eBEER algorithm to the {\it Kepler} light curves, identified a sample of non-eclipsing binary bright candidates, and followed three of them with the Wise observatory eShel spectrograph \citep{Engel2017}. After confirming  the three systems are indeed RV binaries, we fitted the light curves {\it and} the RV data with PHOEBE 
\citep{PHOEBE2.0, PHOEBE2.1}, a detailed numerical light-curve and RV model, estimating the orbital elements and the stellar properties of the two components of each system. Finally, we compared the PHOEBE derived parameters with those obtained by the eBEER approximation and showed that the differences are relatively small.

Section 2 presents the generalized eBEER model, and
Section 3 details the search algorithm for eccentric binaries in the {\it Kepler} data.
Section 4 presents the observations and analysis of three test cases, and
 compares the PHOEBE and eBEER results of these three binaries.
Section 5 summarizes and discusses our results.

\section{The Eccentric BEER Model}

We define the eBEER model for the observed brightness modulation {\it of the primary star}  as
%
\begin{equation}
\mathcal{M_{\rm BEER,1}}(\hat{t}) = \mathcal{M_{\rm beam,1}}(\hat{t}) +  \mathcal{M_{\rm ellip,1}}(\hat{t}) + \mathcal{M_{\rm refl,1}}(\hat{t}) \ ,
\label{eq:mbeer}
\end{equation}
%
where $\mathcal{M_{\rm beam,1}}$, $\mathcal{M_{\rm ellip,1}}$, and $\mathcal{M_{\rm refl,1}}$ are the primary's beaming, ellipsoidal, and reflection (by the  secondary) modulations, respectively, 
and $\hat{t}$, the orbital phase, is the fractional part of $\frac{t-T_0}{P_{\rm orb}}$, where $P_{\rm orb}$ is the orbital period and $T_0$ is the periastron passage time. 

In what follows we consider each of the three modulations.

\subsection{Beaming}

For the beaming modulation, which is proportional to the RV of the primary, we use the known eccentric RV modulation formula to write 
\begin{equation}
\mathcal{M_{\rm beam,1}} =
-2830 \ \alpha_\mathrm{beam,1}
\frac{q}{(1+q)^{2/3}}
\left(\frac{M_1}{M_{\odot}}\right)^{1/3}
\left(\frac{P_\mathrm{orb}}{{\rm day}}\right) ^{-1/3}
 \sin i \ \
\frac{\cos(\omega + \nu)}{\sqrt{(1-e^2)}} \ {\rm ppm}\ ,
\label{eq:beam}
\end{equation}
where 
 $M_1$ and $M_2$ are the masses of the primary and the secondary, respectively, and $q=M_1/M_2$,
$e$ is the eccentricity, $\nu$ is the true anomaly, and $\omega$ is the argument of periastron. 
The time dependence of the beaming effect, as in the other two effects, is presented by $\nu(\hat{t})$. 
The $\alpha_\mathrm{beam,1}$ is  the primary's beaming coefficient of order unity \citep[see, e.g.,][]{Faigler2011b, Faigler2013}.
%

\subsection{Ellipsoidal}

The ellipsoidal model is based on Eq.~1 of \cite{morris93} (MN93) with two modifications. 

\begin{enumerate}
\item In the first modification to the model, we replace any appearance of $a$, the semi-major axis, with 
$\frac{a}{\beta} \equiv \frac{a(1-e^2)}{1+e\cos\nu(\hat{t})}$, the binary separation as a function of the orbital phase.
This is true for all terms of the equation, except for the first term, which is proportional to $(2+5q)/a^3$ and should be treated more carefully. 
This $(2+5q)/a^3$ term represents two different astrophysical effects.
\begin{itemize}
 \item  The first effect describes the deformation of the star due to its self rotation, which is proportional to $P_{rot}^{-2}$, $P_{rot}$ being the star rotation period. If we extend the tidal locking assumption of MN93 of $P_{rot}=P_{orb}$ to eccentric orbits, we get that the rotation term is proportional to $(1+q)/a^3$. 
Note, however, that $P_{rot}$ might be different from $P_{orb}$, as pointed out by \citet{hut81}.
%
 \item The second effect is due to the tidal force from the secondary that is proportional to  $q/r^3$, $r$ being the momentary binary separation.
\end{itemize}
Hence, we need to split the first term in Eq.~1 of MN93 into two parts: 
\begin{itemize}
\item A rotation term which is proportional to $(2+2q)/a^3$ (i.e. to $P^{-2}$). This  term is constant throughout the orbital phase, and does not contribute to the modulation of the stellar brightness.
\item A tidal term which is proportional to $3q / \left(\frac{a}{\beta}\right)^3$, namely a term that varies with phase through the binary momentary separation.
\end{itemize}

 \item In the second modification to the model, we substitute any appearance of the MN93 orbital angle 
$\phi$ with $\omega + \nu(\hat{t}) - \pi /2$.
\end{enumerate}

\newpage 

These two modifications assume an instantaneous response of the stellar surface to the changing tidal force due to the eccentric orbit.
After transforming  $a$ to $P_{\rm orb}$, using Kepler's third law, the ellipsoidal model becomes
%

\begin{alignat}{5}
\label{eq:ellip}
\mathcal{M_\mathrm{ellip,1}} =
13435 \cdot 2\alpha_\mathrm{e0,1}\left(2-3 \sin^2 i\right)
&\left(\frac{M_1}{M_{\odot}}\right)^{-1}
& 
&\left(\frac{P_\mathrm{rot}}{{\rm day}}\right) ^{-2}
&\left(\frac{R_1}{R_{\odot}}\right)^3
&\\
\nonumber
+13435 \cdot 3\alpha_\mathrm{e0,1}\left(2-3 \sin^2 i\right)
&\left(\frac{M_1}{M_{\odot}}\right)^{-1}
&\frac{q}{1+q}
&\left(\frac{P_\mathrm{orb}}{{\rm day}}\right) ^{-2}
&\left(\frac{\beta R_1}{R_{\odot}}\right)^3
&\\
\nonumber
+759 \ \alpha_\mathrm{e0b,1} 
\left(8-40 \sin^2 i +35 \sin^4 i \right)
&\left(\frac{M_1}{M_{\odot}}\right)^{-5/3}
&\frac{q}{(1+q)^{5/3}}
&\left(\frac{P_\mathrm{orb}}{ {\rm day}}\right) ^{-10/3} 
&\left(\frac{\beta R_1}{R_{\odot}}\right)^5
&\\
\nonumber
+3194 \ \alpha_\mathrm{e1,1} 
\left(4 \sin i - 5 \sin^3 i \right)
&\left(\frac{M_1}{M_{\odot}}\right)^{-4/3}
&\frac{q}{(1+q)^{4/3}}
&\left(\frac{P_\mathrm{orb}}{{\rm day}}\right) ^{-8/3} 
&\left(\frac{\beta R_1}{R_{\odot}}\right)^4
&\sin(\omega+\nu) \\
\nonumber
+13435 \ \alpha_\mathrm{e2,1} \sin^2 i
&\left(\frac{M_1}{M_{\odot}}\right)^{-1}
&\frac{q}{1+q}
&\left(\frac{P_\mathrm{orb}}{{\rm day}}\right) ^{-2}
&\left(\frac{\beta R_1}{R_{\odot}}\right)^3
&\cos2(\omega+\nu)\\
\nonumber
+759 \ \alpha_\mathrm{e2b,1} 
\left(6 \sin^2 i - 7 \sin^4 i \right)
&\left(\frac{M_1}{M_{\odot}}\right)^{-5/3}
&\frac{q}{(1+q)^{5/3}}
&\left(\frac{P_\mathrm{orb}}{ {\rm day}}\right) ^{-10/3} 
&\left(\frac{\beta R_1}{R_{\odot}}\right)^5
&\cos2(\omega+\nu) \\
\nonumber
+3194 \ \alpha_\mathrm{e3,1} \sin^3 i
&\left(\frac{M_1}{M_{\odot}}\right)^{-4/3}
&\frac{q}{(1+q)^{4/3}}
&\left(\frac{P_\mathrm{orb}}{{\rm day}}\right) ^{-8/3}
&\left(\frac{\beta R_1}{R_{\odot}}\right)^4
&\sin3(\omega+\nu) \\
\nonumber
+759 \ \alpha_\mathrm{e4,1} \sin^4 i
&\left(\frac{M_1}{M_{\odot}}\right)^{-5/3}
&\frac{q}{(1+q)^{5/3}}
&\left(\frac{P_\mathrm{orb}}{{\rm day}}\right) ^{-10/3}
&\left(\frac{\beta R_1}{R_{\odot}}\right)^5
&\cos4(\omega+\nu)
\ {\rm ppm},
\end{alignat}
%
where 
\begin{equation} 
\beta = \frac{1+e\cos\nu}{1-e^2} \ ,
\end{equation} 
and the $\alpha_{\rm e}$ coefficients are
\begin{equation} 
\alpha_\mathrm{e1,1}=\frac{15u(2+\tau)}{32(3-u)} \ , \ 
\alpha_\mathrm{e2,1}=\frac{3(15+u)(1+\tau)}{20(3-u)} \ , \ 
\alpha_\mathrm{e2b,1}=\frac{15(1-u)(3+\tau)}{64(3-u)} \ ,
\end{equation} 
\begin{equation*}
\alpha_\mathrm{e0,1}=\frac{\alpha_\mathrm{e2,1}}{9} \ , \ \alpha_\mathrm{e0b,1}=\frac{3\alpha_\mathrm{e2b,1}}{20} \ , \
\alpha_\mathrm{e3,1}=\frac{5\alpha_\mathrm{e1,1}}{3} \ , \ \alpha_\mathrm{e4,1}=\frac{7\alpha_\mathrm{e2b,1}}{4} \ ,
\end{equation*} 
%
where $u$ and $\tau$ are the primary's linear limb and gravity darkening coefficients, respectively.  

\subsection{Reflection}

We turn now to  the reflection of the primary light by the secondary surface. For this modulation we use the Fourier series expansion of a Lambert phase function \citep{Faigler2015}. Following the same instantaneous response-based modifications as in the ellipsoidal model we get
%
\begin{align}
\begin{aligned}
\mathcal{M_\mathrm{refl,1}} =
& 56514 \ \alpha_\mathrm{refl,1} 
(1+q)^{-2/3}
\left(\frac{M_1}{M_{\odot}}\right)^{-2/3}
\left(\frac{P_\mathrm{orb}}{ {\rm day}}\right) ^{-4/3}
\left(\frac{\beta R_2}{R_\mathrm{\odot}}\right)^2 
\\
\times
& \big( 0.64 -\sin i\sin\left(\omega+\nu\right) + 0.18 \sin^2 i \left(1- \cos2\left(\omega+\nu\right) \right) \big) \ {\rm ppm} \ .
\end{aligned}
\label{eq:refl}
\end{align}

In the above equations $M_1$, $R_1$, $M_2$, and $R_2$ are the mass and radius of the primary and the secondary, $i$ is the orbital inclination. The $\alpha_\mathrm{beam,1}$, $\alpha_\mathrm{e*,1}$, and $\alpha_\mathrm{refl,1}$ are the primary's beaming, ellipsoidal and reflection coefficients, respectively \citep[for a detailed description of these coefficients see, e.g.,][]{Faigler2011b, Faigler2013}.

\subsection{The full model}

As noted by MN93 the same equations can be used to model the variability of the secondary by interchanging subscript 1 and 2, replacing  $\omega_1$ by $\omega_2=\omega_1+\pi$, and  the primary's $\alpha$ coefficients by the secondary's $\alpha$ coefficients. Consequently, $M_1$ and $M_2$ are interchanged, and $q$ is replaced by $1/q$.

The full eBEER model combines the contribution of the primary and the secondary to produce the total brightness modulation of the system. This is done by 
%
\begin{equation} 
\mathcal{M_{\rm BEER}}(\hat{t}) = f_1 
\mathcal{M_{\rm BEER,1}}(\hat{t})  + f_2 
\mathcal{M_{\rm BEER,2}}(\hat{t}) \ , 
\end{equation}
%
where $f_1$ and $f_2$ are the relative contributions of the primary and the secondary to the total flux in the observed band, respectively, so that by definition $f_1+f_2=1$.

Obviously, our simplistic model for the ellipsoidal and reflection modulations ignores some important astrophysical effects.
The ellipsoidal model neglects additional harmonics of the variability that involve higher orders of $\frac{R_1}{r(\hat{t})}$ \citep{morris93}, and also ignores the time lag due to the response time of the star surface to the changing tidal-force, due to the inherent variable separation and non-synchronization of an eccentric orbit \citep[e.g.,][]{mazeh08}. 
The reflection model ignores, among others, heating time lag due to the changing separation and substellar point in an eccentric orbit, and phase shifts due to heat circulation and other effects. Nevertheless, this model provides manageable approximations of these variations, with minimum additional free parameters, which enabled us to perform an efficient search for eccentric binaries. 

To confront the eBEER approximation with a more detailed model and RV data, we applied the  model to the light curves of {\it Kepler}, detected a sample of non-eclipsing binary candidates and followed three bright systems with RV observations.  
A detailed PHEOBE model was then fitted and compared with the approximated eBEER model.

\section{Search of the {\it Kepler} Light Curves for eccentric binaries}
\label{sec:analysis}
To identify eccentric binaries candidates, we searched the {\it Kepler} Q2--Q16 DR24 raw long-cadence light curves of stars brighter than $10$-th {\it Kepler} mag with  the BEER search algorithm, after adaptation for the Eccentric BEER model.

First, we cleaned the light curves of outliers and jumps and detrended them following \citet{Faigler2013}. Next, we calculated the Fast Fourier Transform (FFT) based power spectrum of the cleaned and detrended light curve, and selected ten candidate orbital periods for each star \citep[See details in][]{Faigler2013}.

 For each period we then derived the amplitudes of its first four harmonics,
 $\{A_i; i=1,8\}$,
 following \citet{Faigler2015b}.
In short, this was done by simultaneously fitting the raw light curve of each quarter to three sets of functions: long-term cosine-detrend functions of periods down to a minimum of twice the orbital period \citep{mazeh10}, cosine and sine functions of the first four orbital-period harmonics, and jump functions at predefined {\it Kepler} times \citep{Faigler2013}.

The uncertainties of the amplitudes were approximated from the quarter-to-quarter scatter of the fitted amplitudes \citep{Faigler2015}.
For each candidate orbital period, the resulting eight amplitudes with their uncertainties, together with the system parameters from the DR24 {\it Kepler} Input Catalog (KIC) \citep{huber14}, were used for fitting to the eBEER model, as described below.

\subsection{Fitting the periodic light curve with the eBEER Model}

At the first stage of the fitting, the model assumes that the secondary is a main sequence star and uses the approximated main sequence relations \citep{salaris05,Duric2004}
\begin{equation}
\frac{L_2}{L_{\odot}} \approx
    \begin{cases}
    0.23 \left(\frac{M_2}{M_{\odot}}\right)^{2.3} &  \ \  M_2 < 0.43 M_{\odot}\\
           \left(\frac{M_2}{M_{\odot}}\right)^{4} &  \ \   0.43 M_{\odot} < M_2 < 2 M_{\odot}\\
    1.5       \left(\frac{M_2}{M_{\odot}}\right)^{3.5} &  \ \   2 M_{\odot} < M_2 < 20 M_{\odot} \ ,\\
    \end{cases}
\label{eq:msapprox1}
\end{equation} 
\begin{equation}
\frac{R_2}{R_{\odot}} \approx \max \left( 0.1 , \left(\frac{M_2}{M_{\odot}}\right)^{0.8} \right) \ ,
\label{eq:msapprox2}
\end{equation}
with the black-body luminosity equation 
\begin{equation}
\left(\frac{L_{x}}{L_{\odot}}\right) = \left(\frac{T_{\rm eff,x}}{T_{\odot}}\right)^{4} \left(\frac{R_{x}}{R_{\odot}}\right)^{2} \ ,
\label{eq:bb}
\end{equation}
where $L_{x}$, $T_{\rm eff,x}$ and $R_{x}$ are the bolometric luminosity, effective temperature and radius of the primary $(x=1)$ or the secondary $(x=2)$, respectively.  These constraints on the secondary are lifted in the last stage of the analysis. 

The light curve model, outlined by Eq.~\ref{eq:mbeer}--\ref{eq:bb}, is a function of  a set of eleven parameters
%
\begin{equation}
\overrightarrow{p}=(P_{\rm orb}, T_0, M_1, R_1, T_{\rm eff,1}, M_2, \sin i, e, \omega, \alpha_\mathrm{refl,1}, \alpha_\mathrm{refl,2})\, ,
\end{equation}
%
while the other six parameters---$R_2$, $T_{\rm eff,2}$, $f_1$, $f_2$, and $\alpha_\mathrm{beam}$ and $\alpha_\mathrm{e*}$ of the primary and the secondary, are all directly estimated from the first set of parameters.

For any set of the eleven parameters, $\overrightarrow{p}$, we derive the expected eight amplitudes
 $\{E_i(\overrightarrow{p}); i=1,8\}$ 
of the eBEER model and compare them to the eight amplitudes of the lightcurve   $\{A_i; i=1,8\}$.
This is done by  minimizing  the $\chi^2$ function,  defined as
%
\begin{equation}
\chi^2(\overrightarrow{p}) = 
\sum \limits_{i=1}^{8} 
\left(\frac{A_i-E_i(\overrightarrow{p})}{\sigma_i} \right)^2 
         +  \sum \limits_{i}  \ln \left(\mathcal{F}_{\theta_i}(\theta_i) \right) \,.
\label{eq:chi2}
\end{equation}
%

The first sum measures the goodness of fit of the data to the model, but does that in the Fourier domain.
As our algorithm is limited to low to medium eccentricity, using the Fourier amplitudes of the first $4$ harmonics, which is more computationally efficient, serves as a good proxy for the time-domain goodness of fit. 

The second sum of Eq.~\ref{eq:chi2} measures the fit of a subset 
of parameters
%
\begin{equation}
\overrightarrow{\theta}=(M_1, R_1, T_{\rm eff,1}, \sin i, \alpha_\mathrm{refl,1}, \alpha_\mathrm{refl,2})\, 
\end{equation}
to their prior-distribution Probability-Density Function (PDF), $\mathcal{F}_{\theta_i}$. 
For the primary's mass and radius the prior distributions are log-normal centered at their KIC values with $1 \sigma$ deviations of $\sim20\%$.
For the orbital inclination, the prior is the distribution for an isotropic orbital-angular-momentum vector.
For each of the $\alpha_\mathrm{refl,x}$ coefficients, the prior is a Gaussian distribution with a mean and standard deviation of $0$ and $0.5$, respectively, with only positive values allowed.

To speed up the calculation of the expected coefficients of the model,
 $\{E_i(\overrightarrow{p}); i=1,8\}$, we prepared for each additive term in Eqs.~\ref{eq:beam}, \ref{eq:ellip} and \ref{eq:refl} a function that depends on $e$, 
$\omega$, 
$T_0$, $M_2$ and $\sin i$ for predetermined points on a 5D grid of these parameters. 
We then derived, using FFT, eight Fourier amplitudes of the first four harmonics of these functions.  
This set of eight coefficients for each grid point and for each term of the model was calculated only once, and was used in the search over the eleven-parameter space for all analyzed stars and possible periods.

At the first stage of the search, for each system we used each of the candidate orbital periods, and set $M_1$,  $R_1$ and $T_{\rm eff, 1}$  to their KIC values.
%
Taking advantage of the linearity of the Fourier transform, we used the system parameters and the previously prepared Fourier amplitudes of the model terms to derive the eight Fourier amplitudes of the full light-curve model 
$\{E_i(\overrightarrow{p}); i=1,8\}$ 
at each grid point. This was done by obtaining   $\alpha_\mathrm{refl,1}$ and $\alpha_\mathrm{refl,2}$ that minimize the $\chi^2$,
using the fact that the model is linear in these two parameters. 
Next, we choose as our first estimate of the system solution the grid point with the lowest $\chi^2$. 

At the second stage of the search we start from the grid point with minimum $\chi^2$,  and use a non-linear least-square fitting algorithm \citep[Trust-Region-Reflective,][]{coleman94,coleman96} to fine-tune $M_1$ and $R_1$ together with the other system parameters.
The result of this fitting is a set of system parameters and their corresponding $\chi^2$ for any given period. 


\subsection{Sorting through the {\it Kepler} star sample}

Finally, at the last stage of the search, we are interested in identifying the most compelling candidates for eccentric binary systems.
To do that we follow \citet{TalOr2015} and assign each candidate orbital period of each star with a score in the $0$--$1$ range, with 1 being the best.
As described in detail in \citet{TalOr2015}, the score is a product of $5$ sub-scores, each of which measures the likelihood of a different aspect  of the system fit to be originating from a real binary.
We then sort all the candidate orbital periods of all stars by their score and visually inspect the top-scoring systems.

To get a better handle on the goodness of fit of our top candidates, we finally directly fit the {\it Kepler} light curve to the time-domain eccentric model, while fine-tuning the system parameters, again using a non-linear least-square fitting algorithm. This fitting is performed for the light curve of each {\it Kepler} quarter separately, and the best-fit parameters are derived as the median of these fitted parameters over the {\it Kepler} quarters. The uncertainty of each parameter is estimated from the scattering of its values over the different {\it Kepler} quarters \citep{Faigler2015}.

\section{Three eccentric binaries as test cases}

Of the top hits, we have identified three  eccentric binary candidates with good eBEER solutions brighter than {\it Kepler} mag of 9.8 that can be observed with our eShel spectrograph (see below), and therefore can be used as test cases of the algorithm.
The full list of the fainter eccentric candidates will be released in a separate publication.
Table \ref{tab:KICtable} gives the {\it Kepler} Input Catalog (KIC)\footnote{https://archive.stsci.edu/kepler/kic.html} and Gaia  \citep{Gaia2016} data for these three candidates, along  with the eBEER period found.


\begin{table}
	\scriptsize
	\centering
		\begin{tabular}{lrrr}
		\hline
		&\astrobj{KIC 6292925} & \astrobj{KIC 5200778} & \astrobj{KIC 4931390}   \\
		\hline
		Name         & BD+41 3418       & TYC 3140-1505-1     & BD+39 3832   \\
		RA (J2000)    & 19:36:57.91      & 19:41:56.47         & 19:35:37.98  \\
		DEC (J2000)   &+41:37:11.3       & +40:18:57.2         & +40:03:27.5  \\
		\hline
		\hline
		{\bf KIC values:} \\
		{\it Kepler} mag   & 9.7              & 9.5                 & 9.3  \\
		$R_{*}\, \rm [R_{\odot}]$  &2.84          & NA                  & 1.46         \\
		$T_{\rm eff}$[K]     &  8159        & NA                   & 6297 \\
		$log\,g$[dex]     & 3.76         & NA                   & 4.181   \\
		$[Fe/H]$[dex]   & 0.087      & NA                   & -0.313 \\
		$ M_1 \rm [M_\odot]$ & 1.7       &    NA              &   1.2    \\
		\hline
		\hline
		{\bf GAIA Data (DR2):} \\
		$\pi\, \rm [mas]$    & $3.03\pm0.3$      & $4.18\pm0.03$        &$6.62\pm0.07$  \\
		$R_{*}\,\rm [R_{\odot}]$   &       NA     & $2.4\pm 0.1$        &  $1.47\pm0.05$          \\
		$T_{\rm eff}$[K]     &   $8160 \pm150 $       &  $5980\pm100 $        & $6560\pm100$  \\
		$RV$[km/s]       & NA                   &  $-16.4 \pm 6.8$        &    $-2.3\pm3.5$       \\
		\hline
		\hline
		{\bf eBEER periods:} \\
		$P$ [d]  &  $13.6136\pm0.00066$      &  $16.354\pm0.036$              & $7.6058\pm0.0023$ \\
		\hline
	\end{tabular}
	\caption {{\small Data for the three candidate systems. 
			{\it Top:} {\it Kepler} Input Catalog (KIC) data ($M_1$ was calculated from $log\, {\rm g}$ and $R_*$), 
			{\it Middle}: Data from the Gaia Data Release 2 \citep{Gaia2018}.
			The last row gives the detected eBEER period.}
	}
	\label{tab:KICtable}
\end{table}

\subsection{Follow-Up Spectroscopic Observations}

Follow-up spectroscopic observations of the three test cases were carried out with the eShel spectrograph on the 1m telescope of the Wise observatory  during three summer campaigns in 2013, 2014 and 2015.
The eShel is an $R \approx 10,000$ fiber-fed echelle spectrograph, spanning a wavelength range of 4500--7500\AA \,\citep{Engel2017}.
The analysis of each set of spectra was based on a template chosen from the PHOENIX library \citep{husser2013} that best fitted the observed spectra. 

Radial velocities, given in the appendix, were derived by cross correlating the spectra with the template chosen, using the UNICOR algorithm \citep{Engel2017}. The RVs were used, with the photometric period as a prior, to obtain the parameters of the Keplerian eccentric orbit.

\subsection{The PHOEBE and the eBEER binary models}

To fit the {\it Kepler} light curve together with the eShel RVs of the three systems we performed two independent fits, using the PHOEBE and our eBEER models. 
As noted above, the main-sequence constrains on the secondary are lifted in this stage. 

PHOEBE (PHysics Of Eclipsing BinariEs)\footnote{http://phoebe-project.org} \citep{PHOEBE2.0} is an open-source code that models the physics of binary systems to produce light curves and radial velocities of these systems.
PHOEBE 2.1 \citep{PHOEBE2.1} includes models for the three BEER effects, making it a suitable tool for modeling light curves at the precision level of {\it Kepler}.

PHOEBE models a light curve by dividing the stellar surface into many  triangular  elements, summing up the flux over the surface elements visible to the observer.
The number of triangles used to model the star surface determines the accuracy of the resulting calculated light curve.
A small number of triangles generates a `noisy' light curve, due to the low sampling of the stellar surfaces. A high number of triangles, on the other hand, results in longer model calculation CPU time.
We found out that with 10,000 triangles per star, a small numerical noise was still present in the light curve, with 
 a negligible impact on its shape, while the calculation time was acceptable.
 
As an initial guess for the PHOEBE fit we used the parameters of the best eBEER models found from the light curves, the orbital solutions of the spectroscopic data, and additional information from KIC and Gaia DR2 data.
We applied an iterative parallel `greedy' search algorithm, 
\citep[e.g.,][]{greedy98}
to find the  minimum $\chi^2$ around the starting point. 
Uncertainties of the fit parameters were estimated by performing  `cuts' in the $\chi^2$ hyper-surface along the parameter axes through the minimum  point found.

We used the same 'greedy' search to identify the best eBEER solution. As eBEER is much faster than PHOEBE, we used the MCMC procedure \citep{foreman13} to refine the solution and find the uncertainties of the eBEER parameters.

\begin{figure}
	\centering
	\resizebox{15cm}{6.5cm}
	{\includegraphics{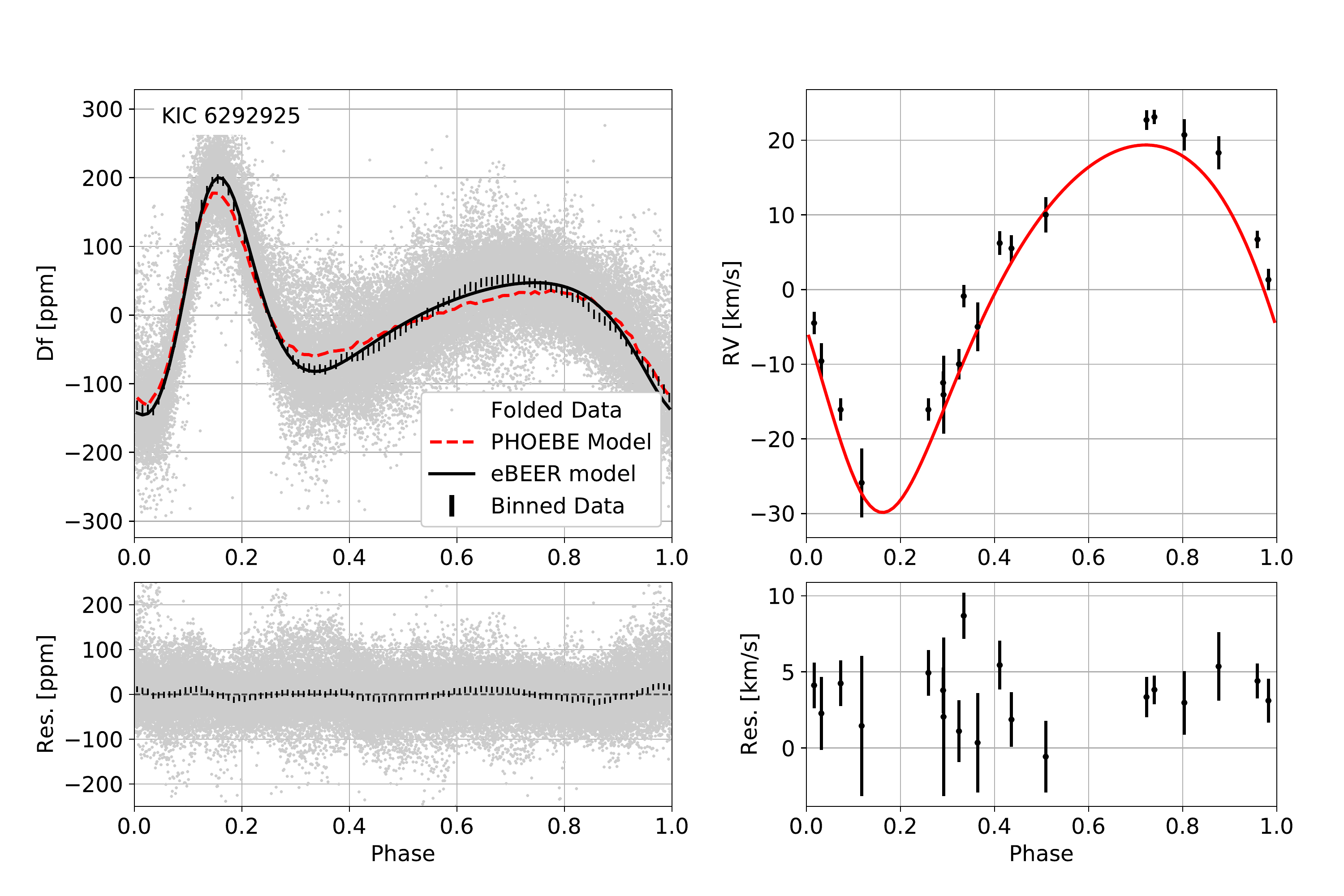}}
	\centering
	\resizebox{15cm}{6.5cm}
	{\includegraphics{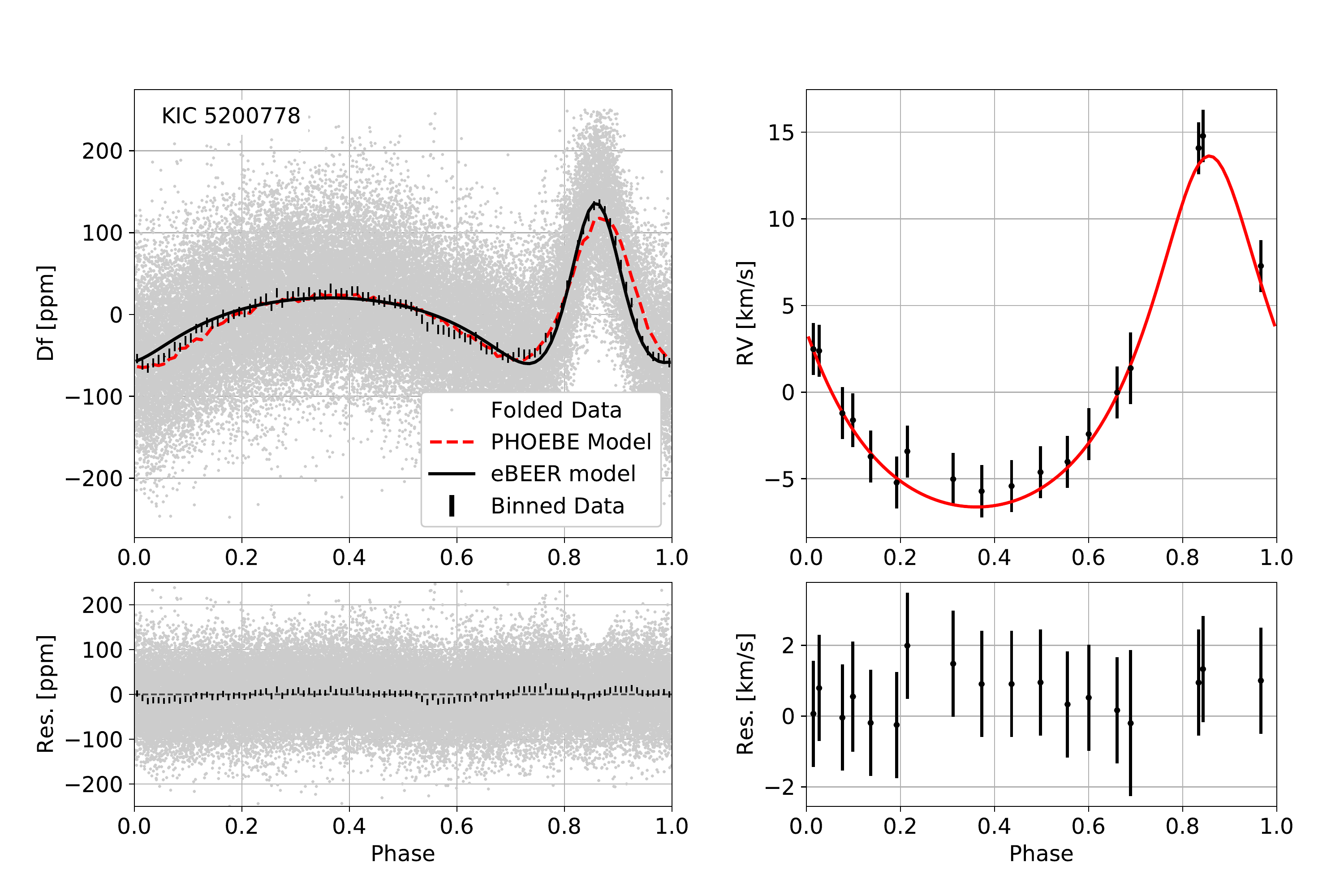}}
	\centering
	\resizebox{15cm}{6.5cm}
	{\includegraphics{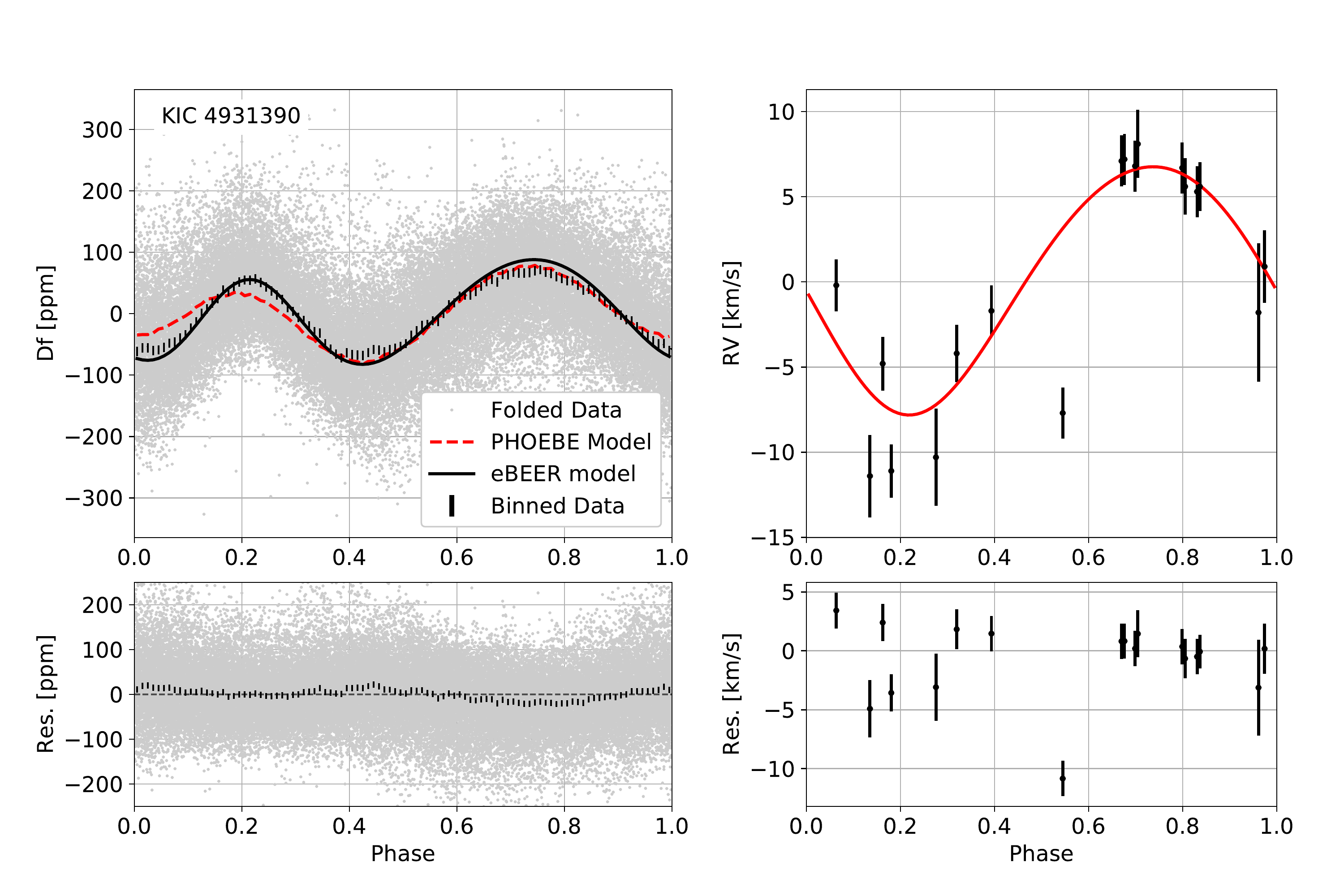}}
	\caption{{\small Folded light curve (left) of the three systems, together with the eBEER (black line) and PHOEBE (dashed red line) models. The light-curve panels include the data binned to 100 bins (vertical bars). The light-curve residuals relative to the eBEER model are shown below each panel.  RV data vs. the PHOEBE model (continuous red line) appear on the right side. 
	}}
	\label{fig:PhasePlots}
\end{figure}

Figure~\ref{fig:PhasePlots} shows the light curves of the three systems, folded with their PHOEBE period, together with the two models. The figure also displays the RVs with the corresponding PHOEBE model.
The figure shows that the eBEER and PHOEBE models fit quite well the folded light curves. The random residuals, reflecting the {\it Kepler} noise per measurement, are on the order of $50$ ppm, while the systematic errors are less than $\sim10$ ppm. 

The orbit parameters of the two models are given in Table \ref{tab:orbit_param_summary} and the stellar parameters in Table~3. Table~2 includes the derived amplitudes of the three BEER effects as found by eBEER. 
One can see that the ellipsoidal effect is significantly detected in the three binaries, while the beaming effect is observed only for KIC 5200778.
Therefore, the stellar radii and temperatures of the secondaries found by eBEER are questionable.  

 All together, the figure and the two tables suggest that the parameters found by eBEER are fitting the light curves quite well and are similar to those of the PHOEBE model, demonstrating the reliability of the eBEER approach.

%
\begin{table}
	\centering
	\begin{tabular}{lrrr}
		\hline
		Parameter      & eBEER Model  &   PHOEBE model\\
		\hline
		\hline
		{\bf KIC 6292925} &   &  & \\
		$P$ [day]      & $13.6126\pm0.0004 $   &     $13.6122 \pm 0.00043$  \\
		$T_{\rm 0}$ [BJD~-~2454833]    & $874.04 \pm 0.03 $   &  $875.977 \pm 0.035$  \\
		$e$            &	$0.232 \pm0.006  $	    &  $0.2491 \pm 0.0085$   \\
		$\omega$ [deg]  &$ 157\pm 2  $ & $153 \pm 3$\\
		$\ i $ [deg]	& $ 71\pm 11 $   &  $ 72\pm 2 $  \\
		$ K$ [km/s]    &           $24.4\pm 0.7   $              &     $24.15 \pm 0.028$     \\
		$ A_{\rm beam} [ppm]$        &	$16\pm10$								&								\\
		$ A_{\rm ellip.} [ppm]$       &	  $180\pm16$ 								&								\\
		$ A_{\rm refl.} [ppm] $ 		&	$22\pm15$								&								\\
		\hline
		{\bf KIC 5200778} &   &  & \\
		$P$ [day]           		& $16.357\pm0.002$    &    $16.3536 \pm 0.0017$     \\
		$T_{\rm 0}$ [BJD-2454833]    &  $806.6\pm0.2 $        &      $ 804.02 \pm 0.13 $  \\
		$e$          				&	$0.35\pm0.03$			&       $ 0.26 \pm 0.011 $   \\
		$\omega$ [deg]    			&  $-1\pm4$ 					&   $2 \pm 4 $  \\
		$\ i $ [deg]				& $23\pm3$  					&  $24 \pm 1 $  \\
		$K$ [km/s]        		 	&  $10.2\pm0.6$            		 &  $ 9.535 \pm 0.015 $   \\
		$ A_{\rm beam} [ppm]$        &	$128\pm9$								&					\\
		$ A_{\rm ellip.} [ppm]$       &	  $183\pm14$ 								&								\\
		$ A_{\rm refl.} [ppm] $ 		&	$7\pm9$								&								\\
		\hline
		{\bf KIC 4931390}  &   &  & \\
		$P$ [day] & $ 7.6064\pm0.0003 $          &         $7.6080\pm 0.0024$                      \\
		$T_{\rm 0}$ [BJD-2454833]      &  $798.44\pm0.04$   & $798.78 \pm 0.12 $  \\
		$e$          &	$0.08\pm0.01 $			&     $0.040 \pm 0.015 $           \\
		$\omega$ [deg] &  $158\pm6 $ 						&  $160 \pm 10$ \\
		$\ i $ [deg]	& $ 50\pm17$ 						 &  $ 35\pm 2$  \\
		$K$ [km/s]     &           $7.2\pm0.5$               &    $7.573 \pm 0.0022 $             \\
		$ A_{\rm beam} [ppm]$        &	$28\pm17$								&								\\
		$ A_{\rm ellip.} [ppm]$       &	  $104\pm32$ 								&								\\
		$ A_{\rm refl.} [ppm] $ 		&	$ 35\pm31$								&								\\
		\hline
	\end{tabular}  
	
	\caption{{ Orbit parameters derived from the eBEER model and PHOEBE model fit.}}
	\label{tab:orbit_param_summary}
\end{table}

\begin{table}
	\centering
	\begin{tabular}{lllllll}
		\hline
		& $M_1\,[\rm M_\odot]$  & $R_1\,[\rm R_\odot]$   & $T_1$[K] & $M_2\,[\rm M_\odot]$ & $R_2\,[\rm R_\odot]$ & $T_2$[K]\\
		\hline
		\hline
		{\bf KIC 6292925 Ph} & $1.89\pm 0.07$ & $1.80\pm0.04 $ &  $8400\pm 200$ & $0.50\pm0.01$ & $1.10\pm0.02$ &  $7000\pm100$\\
		{\bf KIC 6292925 eB} & $1.9\pm 0.1$ & $1.8\pm 0.1$  & $8300\pm 100$ & $0.50\pm0.05$ & $1.3\pm0.2$ & $6000\pm500$ \\
		\hline
		{\bf KIC 5200778 Ph} & $1.20\pm 0.07$ & $2.30\pm0.06 $  &  $6250\pm400$ & $0.37\pm0.03$ & $ 0.47 \pm 0.12$ & $3500 \pm100$ \\
		{\bf KIC 5200778 eB} & $1.2\pm 0.1$ & $2.4\pm 0.1 $ &  $6000\pm100$& $0.40\pm0.07$ & $  0.14\pm0.2 $ & $ 4000\pm2000$\\
		\hline
		{\bf KIC 4931390 Ph} & $1.10\pm0.13$ &  $1.42\pm 0.06 $ &  $6620\pm130$ & $0.13\pm 0.02 $ & $0.66\pm0.04$  & $ 6500\pm 150$\\ 
		{\bf KIC 4931390 eB} & $1.20\pm0.1 $ & $ 1.48\pm0.1$& $6500\pm100$  &$ 0.10\pm0.04  $ &  $0.7\pm0.2$& $ 5500\pm400$\\
		\hline
		\hline
	\end{tabular} 
		\caption{ PHOEBE (Ph) and eBEER (eB) component parameters derived for the three systems. ($T_2$ of KIC 5200778 Ph was set to the main-sequence temperature corresponding to $M_2$ and $R_2$).} 
	\label{tab:PHOEBE_param_summary}
\end{table}

\begin{figure}
	\centering
	\resizebox{18cm}{7.0cm}
	{\includegraphics{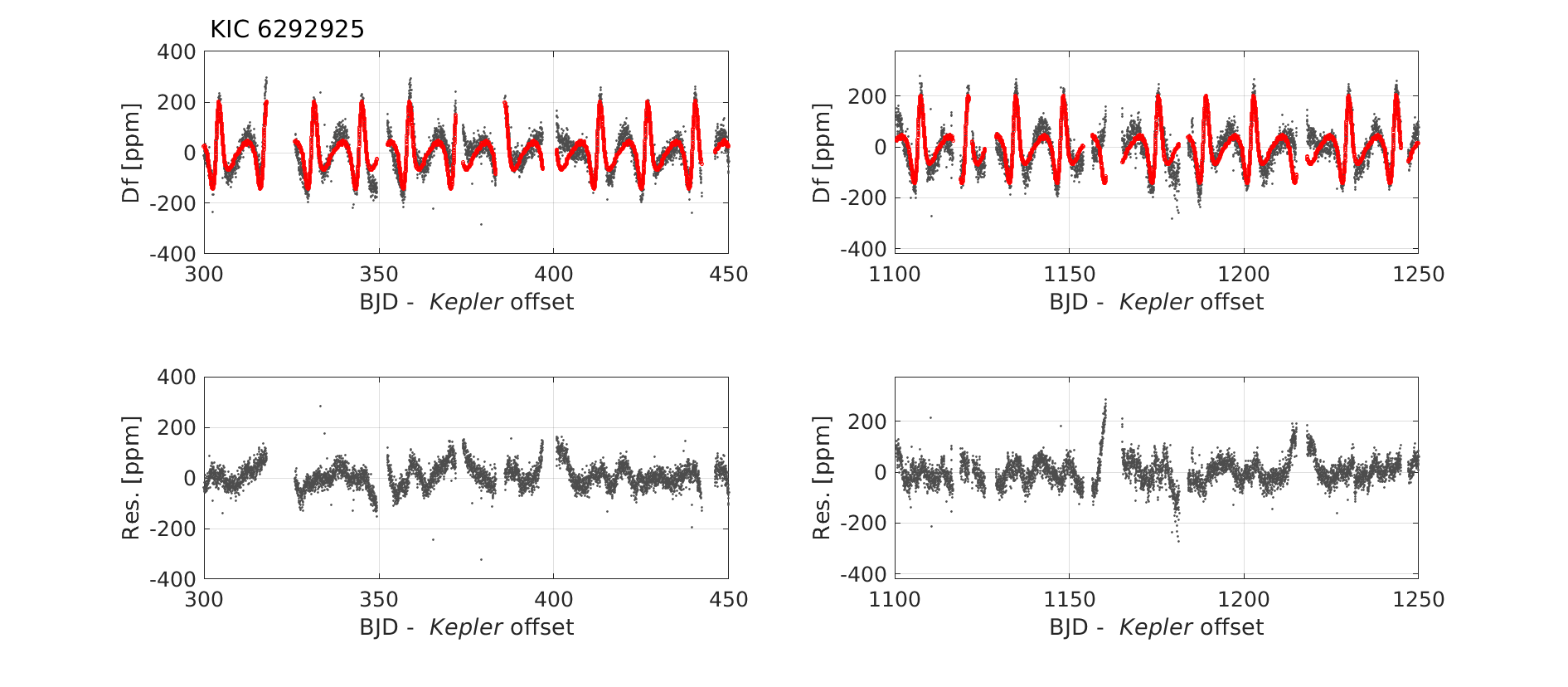}}
	\centering
	\resizebox{18cm}{7.0cm}
	{\includegraphics{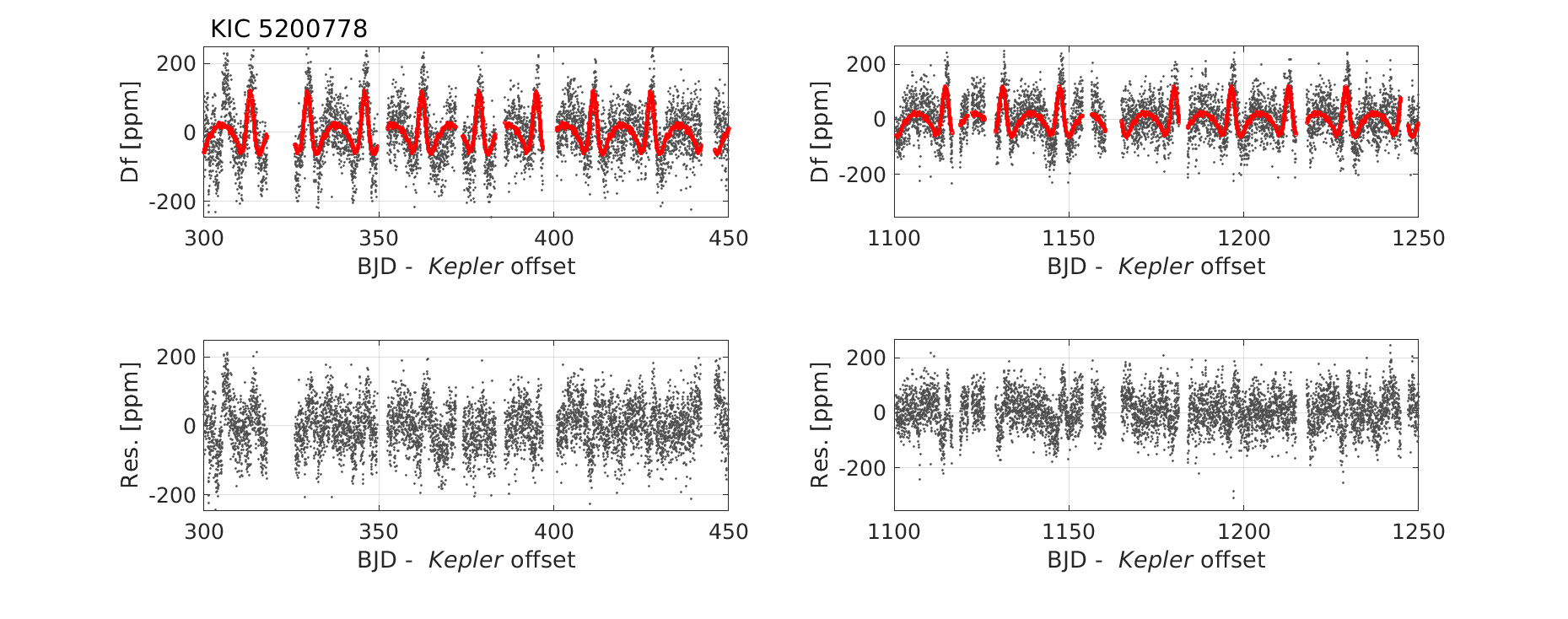}}
	\centering
	\resizebox{18cm}{7.0cm}
	{\includegraphics{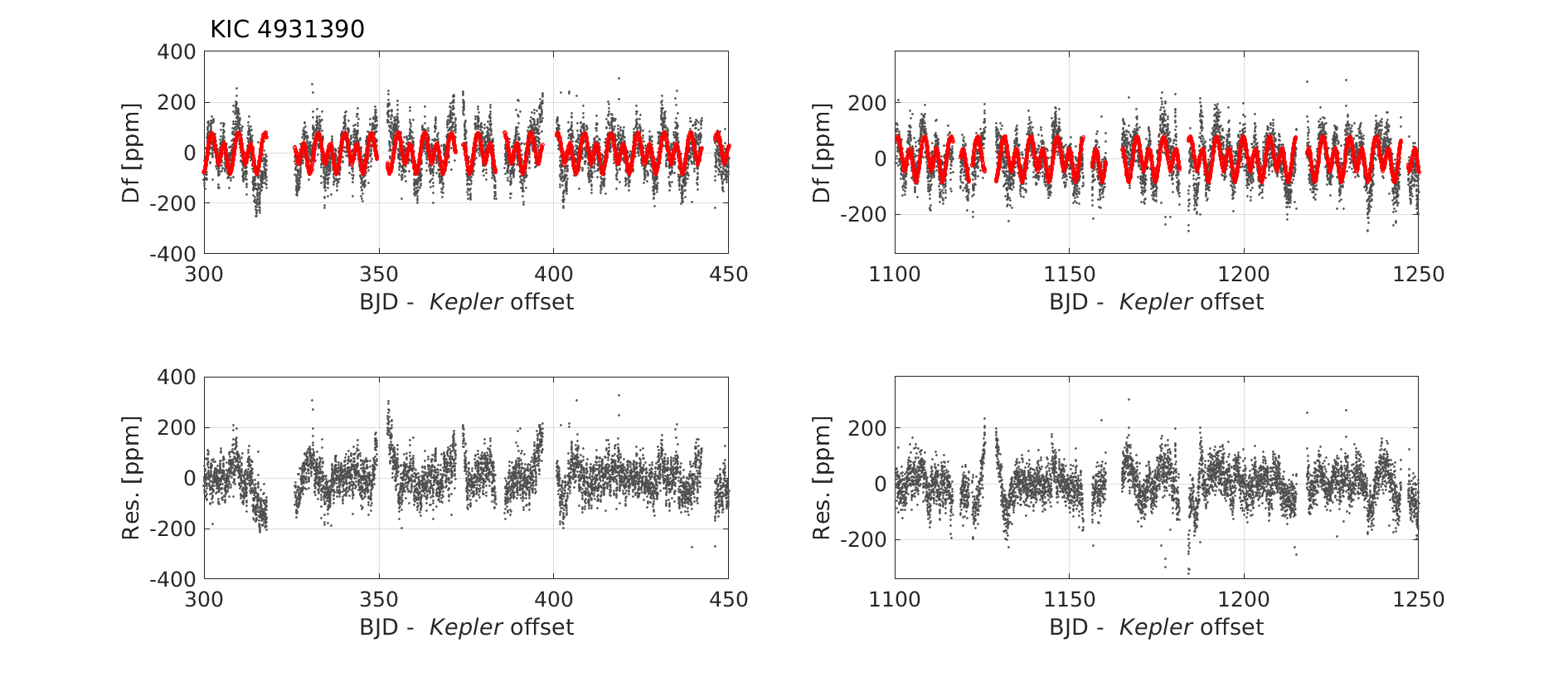}}
	\caption{{\it Kepler} light curves as a function of time for two 150-day sections, together with the PHOEBE model and its residuals.}
	\label{fig:TimePlots}
\end{figure}

Figure~\ref{fig:TimePlots} presents the {\it Kepler} light curves as a function of time for two 150-day sections, 
with the PHOEBE model and its residuals. The figure displays the additional non-periodic modulations hidden in the light curves, which might come from spot modulations. Such variations make the analysis more complicated, and only the long time span of the {\it Kepler} observations enables obtaining the amplitudes of the periodic modulation with high precision. 

\subsubsection{Comments on \astrobj{KIC 6292925} and \astrobj{KIC 4931390}}

Two of the three systems deserve some short comments. 
\astrobj{KIC 6292925} is flagged in the {\it Kepler} eclipsing binaries catalog\footnote{http://keplerebs.villanova.edu} \citep{Kirk2016} as a non-eclipsing "heartbeat" binary system with an orbital period of 13.6 days, consistent with our findings. 

\astrobj{KIC 4931390} was classified  as a variable star with solar-like oscillations by \citet{Molenda2013}.
Periodic variations were detected in several studies and interpreted as spot modulations caused by stellar rotation \citep{Nielsen2013, mcquillan14, Garcia2014, Janes2017}, with periods similar to our orbital period, so we suggest those studies detected the orbital modulation, or that the rotation and orbital periods are synchronized \citep[e.g.,][]{mazeh08}. 

\astrobj{KIC 4931390} displays additional short period variations of $\sim 0.3$ days \citep{Janes2017}.
%
Fitting the modulation with a simple cosine function yielded a period of $0.2926$ day and an amplitude of $56.53$ ppm.  
This modulation was removed from the light curve {\it before} the PHOEBE/eBEER fit. The residuals of this modulation might induce larger errors in the eBEER fitting, reflected in the relatively large uncertainties of the three BEER amplitudes.

\section{Discussion}
%
We present here the eBEER model for the three BEER effects of eccentric binaries, approximated by harmonic series of the true anomaly.
The eBEER model depends on eleven parameters. These include the parameters of the circular BEER model, the stellar parameters---the mass, radius, temperature and luminosity of the two stars, and the orbital parameters of the eccentric orbit,
the eccentricity and the argument of the periastron in particular. 
The approximated model can be used as a tool for fast processing of a multitude of light curves for the detection of eccentric binary systems, and sometimes even eccentric planets. To enhance the efficiency of the search algorithm, we presented here the eBEER model with four harmonics of the mean anomaly, which enables the search to be performed in the Fourier space.

The eBEER introduces a few approximations. 
The beaming model assumes the star is a point source and neglects the stellar rotation and its finite dimensions.
The ellipsoidal model neglects additional harmonics of the variability that involve higher orders of ${R_1}/{r(\hat{t})}$ \citep{morris93}, and also ignores the response time lag of the stellar surface to the changing tidal force, due to the inherent variable separation and angular velocity of an eccentric orbit \citep[e.g.,][]{mazeh08}. 
The reflection model ignores  the heating time lag, due to the changing separation and substellar point in an eccentric orbit, and phase shifts due to heat circulation and other stellar surface effects \citep[e.g.,][]{showman02,knutson07,Faigler2015}. 
Another approximation of eBEER is the expansion of the three modulations to four orbital harmonics only. Because of these approximations, eBEER is not accurate, for highly eccentric binaries in particular. 

Despite its limitations, the eBEER model provides manageable approximations of these variations, with minimum additional free parameters, which enable us to perform an efficient search in hundreds
of thousands of light curves for eccentric binaries. 
When we encounter a suspicious system, either with a large discrepancy between the model and the folded light curve or with high eccentricity, we should extend the model to a higher number of harmonics, or move to a numerical model of the BEER effects.

To verify the algorithm, we searched the light curves of the {\it Kepler} mission for bright eccentric  binary systems, 
observed three of them---KIC 6292925, KIC 5200778 and KIC 4931390, with the eShel spectrograph, and verified that these are indeed RV binaries. We then derived PHOEBE and eBEER solutions for each of the three test cases, based on the {\it Kepler} light curves and the RV measurements. 

As can be seen in Figure~\ref{fig:PhasePlots}, the PHOEBE model describes quite well the stellar variation. The residuals are characterized by a stellar noise of $\sim50$ ppm and a systematic deviation of $\sim10$--$20$
ppm at certain phases. Given the approximations used by PHOEBE, the systematic residuals are relatively small. The  eBEER results are similar to the PHOEBE model, except  for KIC 4931390, which displays a slight difference, as seen in Figure~\ref{fig:PhasePlots}, that gets up to $\sim40$ ppm. This is probably because of  the relatively small orbital separation of the binary, which causes higher orders of ${R_1}/{r(\hat{t})}$ to be non-negligible. 
The similarity between the results of the two models, reflected in 
Tables~\ref{tab:orbit_param_summary} and 3 and Figure~\ref{fig:PhasePlots}, shows that the eBEER and PHOEBE {\it orbital} solutions are quite similar, except for a small eccentricity difference for KIC 4931390. 

Our conclusion is therefore that the eBEER model can be used as a search engine for finding non-eclipsing eccentric binary systems in large photometric databases. The advantage of eBEER over BEER is enhanced for binaries with intermediate periods, at the range of, say, $\sim5$--$20$ days, for which the tidal interaction is not strong enough to circularize the binaries, on one hand  \citep[e.g.,][]{mazeh08}, and the three BEER effects are still detectable on the other hand. This advantage is demonstrated by the three test case binaries studied here. In this range of periods, the number of {\it eclipsing} binaries is decreasing because of the geometrical factor, so finding non-eclipsing binaries is even more important for the study of the binary population.

The shape of the three effects---the beaming, ellipsoidal and reflection modulations, depend on the orbital parameters of the binary, but their relative amplitudes depend also on the stellar parameters of the two stars, 
those of the primary in particular. 
Therefore, modeling the binary light curve is an 
opportunity to obtain some constraints on the stellar parameters of the primary.
However, not like in the eclipsing-binary case, these parameters are weakly coded into the light curves of the ellipsoidal variables. We could see this in the three examples presented here, where the stellar parameters are poorly constrained.  

In order to test the weak dependence of the model on the stellar parameters one could apply the eBEER analysis to eclipsing or  double-lined spectroscopic binaries, for which the radii or mass ratios of the systems are better constrained. However, this is out of the scope of this paper. 

To improve the eBEER analysis,
one could add external information about the system into the model. This includes the system brightness in many photometric bands, from UV to the infrared, and the Gaia DR2 parallax of the system, both available now  for most  bright stars. One can therefore independently derive the temperature and radius of the primary, and sometimes even that of the secondary.
This can serve as prior information to the PHOEBE/eBEER analysis, which models the brightness modulation of a single band, as was done here by introducing some priors on the stellar parameters. 

The prior information could be a crucial component of the analysis.
As we all know, a large number of stars present periodic modulation caused by stellar oscillations, including pulsation and rotation \citep[e.g.,][]{mcquillan14,clementini19,eyer19,eyer19a}. 
The many parameters of the eBEER model allow the search engine to find false models for many different modulations. The prior information on the system can filter out 
such false-positive cases.

This external information is not incorporated into the eBEER search engine yet.  At this stage this approach can be reversed---we can compare the masses and radii found by the light-curve fit with values found by other studies. For example, 
\astrobj{KIC 6292925} was classified  as a late A star \citep{Niemczura2015}, consistent with our findings. On the other hand, the secondary mass of KIC 6292925 was found by the PHOEBE fit as $M_2\sim0.5\,M_\odot$, a value that does not correspond to a main-sequence star with the radius of $R_2\sim1.1\,R_\odot$ and temperature of  $T_{\rm eff,2}\sim7000$ K. 
This  might indicate that eBEER is not that sensitive to the radius and temperature of the secondary, as pointed out above.

The combination of the eBEER search engine, together with the prior available information about the observed systems, can be a powerful tool for finding many thousands of non-eclipsing binaries in current and future photometric surveys \citep{shporer17}, such as OGLE \citep{udalski15}, WASP \citep{pollacco06}, Gaia \citep{Gaia2016}, TESS \citep{ricker15}, Pan-STARRS \citep{kaiser10} and LSST \citep{ivezic19}. Analysis of these systems can be used to deepen our understanding of the population of close binaries. 

\section*{Acknowledgments}

We are indebted to the anonymous referee for pointing our attention to the additional ellipsoidal and reflection equations terms that produce brightness modulations in eccentric orbits.
We deeply thank the PHOEBE team, especially Andrej Pr{\v s}a and Kyle Conroy, for their endless effort to help us run PHOEBE. We could not have completed this work without their
prompt and kind support. 
We thank Ilya Kull for developing the eBEER search-engine code.

This paper includes data collected by the Kepler mission. Funding for the Kepler mission is provided by the NASA Science Mission directorate.

This work has made use of data from the European Space Agency (ESA) mission
{\it Gaia} (\url{https://www.cosmos.esa.int/gaia}), processed by the {\it Gaia}
Data Processing and Analysis Consortium (DPAC,
\url{https://www.cosmos.esa.int/web/gaia/dpac/consortium}). Funding for the DPAC
has been provided by national institutions, in particular the institutions
participating in the {\it Gaia} Multilateral Agreement.

This research was supported by Grant No. 2016069 of the United States-Israel Binational Science Foundation (BSF) and by the Israeli Centers for Research Excellence (I-CORE, grant No. 1829/12).

\section*{Data Availability }
The {\it Kepler} light curve data used in this paper are publicly available in the MAST database (\url{http://archive.stsci.edu/kepler/data_search/search.php}) and the RV data measured for this paper are given in the appendix below.







\clearpage
\appendix
\section{RV Data}

\FloatBarrier
\begin{table}
	\label{tab:K04931390RV}
	\centering
	\scriptsize
	\caption{RV values for the three test cases. Errors smaller than $1.5$ km/s were clamped to this value in the fitting process.}
	
	\begin{tabular}{rrr|rrr|rrr}
		\\
		\hline
		\hline
		\multicolumn{3}{c}{KIC 6292925} & \multicolumn{3}{c}{KIC 5200778}  & \multicolumn{3}{c}{KIC 04931390}\\
		HJD\quad \quad \ & RV\   & $\sigma_{\textsc v}\ $ &
		HJD \quad \quad \  & RV\ & $\sigma_{\textsc v}\ $  & 
		HJD \quad \quad \ & RV \ & $\sigma_{\textsc v}\ $  \\
		& km/s &  km/s      &     & km/s &  km/s       &     & km/s &  km/s       \\
		\hline 
		\hline
		 2456582.21 & -36.6 & 5.2 & 2456583.245 & -29.3 & 2.1 & 2456583.279 & -10.7 & 2.4 \\
        2456583.196 & -27.5 & 3.3 & 2456753.515 & -32.3 & 1.6 & 2456584.346 & -9.6 & 2.9 \\
        2456584.176 & -17.0 & 1.8 & 2456765.534 & -16.6 & 0.8 & 2456585.243 & -1.0 & 1.2 \\
        2456585.168 & -12.5 & 2.4 & 2456768.496 & -28.2 & 1.4 & 2456591.23 & -10.4 & 1.6 \\
        2456596.27 & -32.5 & 2.0 & 2456769.503 & -31.9 & 0.7 & 2456592.288 & -3.5 & 1.7 \\
        2456745.549 & -35.0 & 0.6 & 2456770.488 & -34.4 & 1.3 & 2456595.173 & 7.5 & 0.7 \\
        2456752.53 & -1.8 & 2.1 & 2456794.427 & -33.1 & 0.2 & 2456595.217 & 8.8 & 2.0 \\
        2456770.417 & -48.4 & 4.6 & 2456795.415 & -30.7 & 0.7 & 2456596.175 & 6.0 & 1.3 \\
        2456792.501 & 0.6 & 0.9 & 2456820.462 & -35.9 & 0.9 & 2456596.22 & 6.3 & 1.4 \\
        2456794.359 & -4.2 & 2.2 & 2456822.428 & -35.7 & 0.7 & 2456597.168 & -1.1 & 4.1 \\
        2456795.485 & -15.8 & 1.2 & 2456823.417 & -36.4 & 0.6 & 2456597.267 & 1.6 & 2.1 \\
        2456796.476 & -32.1 & 2.4 & 2456824.459 & -36.1 & 0.7 & 2456610.166 & 7.8 & 0.8 \\
        2456819.497 & 0.2 & 1.3 & 2456825.465 & -35.3 & 0.6 & 2456610.214 & 7.9 & 1.1 \\
        2456823.491 & -27.0 & 0.3 & 2456847.471 & -15.9 & 0.7 & 2456611.148 & 7.4 & 1.2 \\
        2456851.486 & -38.6 & 0.6 & 2456849.485 & -23.4 & 0.9 & 2456611.194 & 6.3 & 1.7 \\
        2456877.476 & -21.2 & 1.4 & 2456850.481 & -28.3 & 0.2 & 2457260.459 & -4.1 & 1.6 \\
        2457262.402 & -38.6 & 0.5 & 2456875.476 & -34.7 & 1.1 & 2457263.37 & -7.0 & 1.2 \\
        2457263.425 & -23.4 & 1.5 & 2457262.472 & -34.1 & 0.7 & 2457267.315 & 0.5 & 1.5 \\
        2457264.466 & -16.3 & 1.6 &  &  &  &  &  &  \\
		\hline\hline
	\end{tabular}
	
\end{table}
\FloatBarrier


\bsp	
\label{lastpage}
\end{document}